\documentclass[pra,twocolumn]{revtex4}

\usepackage{graphicx}
\usepackage{dcolumn}
\usepackage{bm}

\newcommand{\bo}{\raise-1mm\hbox{\Large$\Box$}}

\newcommand{\rr}{\mathbf{r}}
\newcommand{\kk}{\mathbf{k}}
\usepackage{epsfig}
\usepackage{latexsym}
\usepackage{mathrsfs}
\usepackage{setspace}
\usepackage{amsmath}
\usepackage{floatflt}
\usepackage{wrapfig}
\long\def\symbolfootnote[#1]#2{\begingroup
\def\thefootnote{\fnsymbol{footnote}}\footnote[#1]{#2}\endgroup}

\begin{document}

\title{Stability and Excitations of a Dipolar Bose-Einstein Condensate with a Vortex}

\author{Ryan M. Wilson\email{rmw@colorado.edu}}
\affiliation{JILA and Department of Physics, University of Colorado, Boulder, Colorado 80309-0440, USA}
 
\author{Shai Ronen}
\affiliation{JILA and Department of Physics, University of Colorado, Boulder, Colorado 80309-0440, USA}

\author{John L. Bohn}
\affiliation{JILA and Department of Physics, University of Colorado, Boulder, Colorado 80309-0440, USA}

\date{\today}

\begin{abstract}
We study the stability of singly- and doubly-quantized vortex states of harmonically trapped dipolar Bose-Einstein Condensates (BECs) by calculating the low-lying excitations of these condensates.  We map the dynamical stability of these vortices as functions of the dipole-dipole interaction strength and trap geometry by finding where their excitations have purely real energy eigenvalues.  In contrast to BECs with purely contact interactions, we find that dipolar BECs in singly-quantized vortex states go unstable to modes with an increasing number of angular and radial nodes for more oblate trap aspect ratios, corresponding to \emph{local} collapse that occurs on a characteristic length scale.  Additionally, we find that dipolar BECs in doubly-quantized vortex states are unstable to decay into a different topological state (with two singly-quantized vortices) for all interaction strengths when the trap geometry is sufficiently prolate to make the dipoles attractive, and in windows of interaction strength when the trap geometry is sufficiently oblate to make the dipoles repulsive.

\end{abstract}

\maketitle

\section{Introduction}

The observation of strong dipolar effects in a Bose-Einstein condensate (BEC) of harmonically trapped $^{52}\mathrm{Cr}$~\cite{KochNature08a,Lahaye08} atoms marks encouraging progress towards understanding these novel systems.  In contrast to the isotropic contact interaction present in condensates of alkali atoms, the dipole-dipole interaction is long ranged, anisotropic, and is predicted to induce interesting ground state structures~\cite{Ronen07,Baranov08PR} and excitation spectra in both fully and partially trapped systems.  For example, an excitation spectrum much like the roton-maxon spectrum in superfluid $\mathrm{He}$ is predicted in dipolar BECs (DBECs) with both three-dimensional (3D)~\cite{Ronen07,Wilson08} and quasi-two-dimensional (2D)~\cite{Santos03a} geometries.  Additionally, the presence of dipolar effects has recently been shown to be critical in explaining the rich behaviors of spinor BEC systems~\cite{Kawaguchi06,Vengalattore08}, the effects of which are strongly conditioned by the attractive part of the dipole-dipole interaction.  For this reason, it is instructive to compare the dipolar system to a BEC with attractive contact interactions.  In this article, we find the properties of the two systems to be in stark contrast.

To see this contrast, we consider the effects of the dipole-dipole interaction on a condensate with a single vortex core~\cite{Pu06}.  The conditions for the generation of such a DBEC vortex state are studied in Ref.~\cite{Bijnen07}.  First, consider a trapped BEC with attractive contact interactions, characterized by the $s-$wave scattering length $a_s$ of the constituent particles and with \emph{no} vortex.  For such a system, there always exists a critical particle number above which the condensate goes unstable, with preference to collapse in the region of maximum density at the center of the trap~\cite{Dodd96,Bradley97}.  Stirring the condensate into a vortex state serves to stabilize the system by introducing a kinetic energy component due to angular momemtum that offsets the interparticle attraction.  So, in general, the vortex will sustain a larger number of particles than the non-vortex state, and is more stable.

The case for a DBEC, however, is quite different.  Consider a DBEC with its dipolar entities polarized in the $z$-direction.  Because the dipole-dipole interaction is anisotropic, the structure and stability of a DBEC depends strongly on the trap aspect ratio $\lambda = \omega_z / \omega_\rho$, where $\omega_z$ and $\omega_\rho$ are the axial and radial trap frequencies, respectively.  For smaller $\lambda$, the dipole-dipole interaction distends the condensate into a prolate shape, where macroscopic collapse can occur due to long-range attraction in the direction of polarization.  As $\lambda$ is increased, the condensate is stabilized since the dipole-dipole interaction is predominitely repulsive in this more oblate geometry.  For a moderate number of particles, the energy cost of stacking the dipoles in the direction of polarization is outweighed by the tight harmonic confinement in this direction.  However, for a sufficiently large number of particles, a DBEC in an oblate trap is subject to an instability due to local density fluctuations, which are foreign to the contact potential case.  The attractive part of the dipole-dipole interaction dominates in regions where the higher-density fluctuations occur, inititating a local collapse of the condensate.  As will be discussed below, this instability is intimitely connected with an excitation that goes soft at a critical number of dipoles, and that has been dubbed the ``discrete-roton.'' ~\cite{Ronen07}  Signatures of this local collapse have been articulated in the simulations of Ref.~\cite{Parker08}, where the manifestation of the roton in the collapse of a DBEC is shown for the early stages of collapse.

In the presence of a singly-quantized vortex, the region of high density is forced away from the center of the trap due to the zero-density of the vortex core.  Depending on the aspect ratio $\lambda$ of the trap, this either serves to stabilize (for smaller $\lambda$) or destabilize (for larger $\lambda$) the DBEC.  For smaller $\lambda$, the vortex core simply breaks the prolate shape of the condensate along the direction of polarization, eliminating much of the attractive dipole-dipole interaction in this direction and thus increasing the energy due to interactions.  Conversely, for larger $\lambda$ the vortex increases the density in the periphery of the core and thus encourages local collapse.  Just as the roton wavelength is set by the the confinement length in the direction of polarization ($z$), the local density fluctuations occur at the same length scale regardless of the trap geometry.  Widening the trap radially while keeping the axial trapping frequency fixed makes more room for regions of density fluctuations instead of enlarging the existing regions.  This marks a clear and important distinction between the dipole-dipole and contact interactions.  An additional consideration relevant to the stability of DBECs with a vortex are the excitations of the vortex core itself~\cite{Klawunn08}.  As we will see below, these excitations are unlikely to play a role in destabilizing the vortex for oblate, or even mildly prolate, traps.

Our paper is organized as follows:  In section~\ref{sec:methods} we describe the model that we use to study the ground state with a vortex, including the novel algorithm that we employ to carry out our calculations.  In section~\ref{sec:BdG} we discuss the Bogoliubov de Gennes (BdG) formalism and present our calculations of the BdG spectrum in reference to the stability of the system.  Finally, in section~\ref{sec:s2} we calculate the BdG spectrum of a DBEC with a doubly-quantized vortex.  Interestingly, in a more prolate trap geometry where the dipoles are mostly attractive we find that the BdG spectrum looks very similar to the BEC with purely attractive contact interactions, while in a more oblate trap geometry where the dipoles are mostly repulsive we find that the BdG spectrum looks very similar to that of the BEC with purely repulsive contact interactions, having windows of dynamical stability for certain dipole-dipole interaction strengths~\cite{Lundh06,Pu99}.

\section{Methods}
\label{sec:methods}

At ultracold temperatures, a condensate of $N$ bosons trapped in an external potential $U(\rr)$ may be described within mean field theory~\cite{Bortolotti06} by the nonlocal Gross-Pitaevskii Equation (GPE):
\begin{eqnarray}
\label{GPE}
\left[ -\frac{\hbar^2}{2M} \nabla^2 + U(\rr) + (N-1) \right. \nonumber \\
\left. \times \int d\rr^\prime \Psi_0^*(\rr^\prime) V(\rr-\rr^\prime) \Psi_0(\rr^\prime)\right] \Psi_0(\rr) = \mu \Psi_0(\rr)
\end{eqnarray}
where $\Psi_0(\rr)$ is the condensate wave function with unit norm, $\rr$ is the distance from the trap center, $M$ is the boson's mass and $V(\rr - \rr^\prime)$ is the two-particle interaction potential.  For a cylindrically symmetric harmonic trap, the external potential is given by $U(\rr) = \frac{1}{2}M\omega_\rho^2(\rho^2+\lambda^2 z^2)$, where $\lambda = \omega_z/\omega_\rho$ is the trap aspect ratio.  The interaction potential for a dipolar system has the form~\cite{Yi00}
\begin{equation}
\label{intpot}
V(\rr-\rr^\prime) = \frac{4\pi\hbar^2 a_s}{M} \delta(\rr - \rr^\prime) + d^2\frac{1-3\cos^2{\theta}}{|\rr-\rr^\prime|^3}
\end{equation}
where $d$ is the dipole moment and $\theta$ is the angle between the vector $\rr-\rr^\prime$ and the dipole axis.  To illuminate purely dipolar effects, we set $a_s=0$ in this work, a limit that has been achieved experimentally in a BEC of atomic $^{52}\mathrm{Cr}$ by employing a Fano-Feshbach resonance~\cite{Werner05}.

To characterize the structure and stability of a DBEC with a vortex, we introduce the dimensionless characteristic dipole strength,
\begin{equation}
\label{D}
D = (N-1)\frac{Md^2}{\hbar^2 a_\mathrm{ho}}
\end{equation}
where $a_\mathrm{ho} = \sqrt{\hbar/M\omega_\rho}$ is the radial harmonic oscillator length.  Notice that increasing (decreasing) $D$ corresponds to either increasing (decreasing) the number of particles in the condensate or the square of the dipole moment of the particles.  So for a DBEC of, say, $^{52}\mathrm{Cr}$, one must change the number of atoms in the condensate in order to change $D$, since the magnetic dipole moment of $^{52}\mathrm{Cr}$ (6 $\mu_B$) is effectively fixed.

The second term in Eq.~(\ref{intpot}) describes the two-body dipole-dipole interaction for dipoles that are polarized along the trap axis ($z$-axis)~\cite{JacksonEM}, as may be achieved by applying a strong external field to the condensate.  This term is long ranged ($\propto 1/r^3$), anisotropic, and has a coordinate-space singularity as the distance between the particles goes to zero.  These concerns are handled by treating the mean field dipole term in the GPE, $d^2\int d\rr^\prime \Psi_0^*(\rr^\prime)\frac{1-3\cos^2{\theta}}{|\rr-\rr^\prime|^3}\Psi_0(\rr^\prime)$, by means of convolution.  This method eliminates the singularity in coordinate space since the dipolar mean field in momentum space, $\tilde{V}_D(\kk)$, is given by~\cite{Goral03}
\begin{equation}
\label{Vk}
\tilde{V}_D(\kk) = \frac{4\pi}{3}(3 k_z^2/k^2-1).
\end{equation}
The dipolar mean field in coordinate space may then be calculated in terms of $\tilde{V}_D(\kk)$,
\begin{equation}
\label{Vr}
\Phi_D(\rr) = \mathcal{F}^{-1}\left[\tilde{V}_D(\kk)\tilde{n}(\kk)\right],
\end{equation}
where $\mathcal{F}^{-1}$ is the inverse Fourier transform operator and $\tilde{n}(\kk)$ is the Fourier transform of the condensate density.

In general, these transforms must be computed in three dimensions to capture the three-dimensional (3D) nature of the system.  However, the system that we are considering possesses cylindrical symmetry in both the trapping and dipolar interaction potentials.   With such a symmetry, the condensate states may be written in cylindrical coordinates as eigenstates of the orbital angular momentum projection $s$, $\Psi_0(\rr) = \psi_0(\rho,z)e^{is\varphi}$~\cite{BEC2003}.  The $s=0$ state corresponds to a rotationless condensate, while the $s>0$ states correspond to condensates with vortices of charge $s$, or equivalently, with $\hbar s$ units of orbital angular momentum per particle.  This formulation allows for a distinctive computational algorithm to be applied to the problem, reducing a fully 3D calculation to a 2D one by working in cylindrical coordinates and integrating out the simple $\varphi$ dependence of the state.  Specifically, the algorithm uses a one-dimensional (1D) Hankel transform of order $s$ in the $\rho$-coordinate and a 1D Fourier transform in the $z$-coordinate to transform a function with the angular dependence $e^{is\varphi}$ into momentum space.  For example, the transform of the wave function $\psi_0(\rho,z)e^{is\varphi}$ is
\begin{eqnarray}
\label{transform}
\tilde{\psi_0}(\kk) = \sqrt{2\pi} i^{-s} e^{-isk_\varphi} \nonumber \\
\times \int_{-\infty}^\infty dz \, e^{-ik_z z} \int_0^\infty \rho \, d\rho \, \psi_0(\rho,z) J_s(k_\rho \rho)
\end{eqnarray}
where $J_s(k_\rho \rho)$ is the Bessel function of order $s$.  In practice, the Fourier transform and the Hankel transform of order $s$ are performed on discrete grids.  The application of this algorithm to solving the GPE is detailed in Ref.~\cite{Ronen06a}.

In addition to calculating the ground state with a specified vorticity using this algorithm, we also extended it to calculate the BdG excitation spectrum in the presence of a vortex ground state (see section~\ref{sec:BdG}), which describes the low-lying excitations and reveals critical information regarding the dynamical stability of the system.  This extension of our algorithm is described in Appendix~\ref{app:numerics}.

\section{Excited States and Stability}
\label{sec:BdG}

To study the elementary excitation spectrum of a DBEC with a given projection of angular momentum $s$ in the ground state, we use the BdG \emph{ansatz} and write a wave function of the time-dependent GPE of the form (with $\hbar = 1$)
\begin{equation}
\label{BdGansatz}
\psi(\rr,t) = \left[ \psi_0(\rho,z) + \vartheta(\rr,t) \right]e^{i(s\varphi -\mu t)}
\end{equation}
where $\mu$ is the chemical potential of the ground state, $\psi_0(\rho,z)e^{is\varphi}$, and $\vartheta(\rr,t)$ is given by
\begin{equation}
\label{vartheta}
\vartheta(\rr,t) = \delta\left[ u(\rho,z)e^{i(m\varphi -\omega t)} + v^*(\rho,z)e^{-i(m\varphi - \omega t)} \right]
\end{equation}
where $\delta \ll 1$ to ensure that the excitations have small amplitudes, $\omega$ is the frequency of the excitation and the modes $u(\rho,z)e^{im\varphi}$ and $v^*(\rho,z)e^{-im\varphi}$ are normalized by~\cite{BEC2003}
\begin{equation}
\label{uvnorm}
\int d\rr \left[ |u(\rho,z)|^2 - |v(\rho,z)|^2 \right] = 1.
\end{equation}
The BdG modes are characterized by the quantum number $m$, being their projection of orbital angular momentum onto the $z$-axis.  This \emph{ansatz} represents a vortex with angular momentum $\hbar s$ per particle giving rise to excitations with angular momentum $\hbar (s \pm m)$.  By inserting Eq.~(\ref{BdGansatz}) into the time-dependent GPE (Eq.~(\ref{GPE}) with $\mu$ on the right hand side replaced by $i\hbar \frac{\partial}{\partial t}$) and linearizing about $\delta$, the coupled BdG equations are derived by collecting terms evolving in time like $e^{-i\omega t}$ and $e^{i\omega t}$, respectively,

\begin{eqnarray}
\label{BdG1}
\omega \, u(\rho,z)e^{im\varphi} = \left[ H_0 - \mu \right. \nonumber \\
\left. + \int d\rr^\prime \psi^*_0(\rho^\prime,z^\prime) V_N(\rr-\rr^\prime) \psi_0(\rho^\prime,z^\prime)\right] u(\rho,z)e^{im\varphi} \nonumber \\
+ \int d\rr^\prime \psi_0^*(\rho^\prime,z^\prime) V_N(\rr-\rr^\prime) u(\rho^\prime,z^\prime)e^{im\varphi^\prime}\psi_0(\rho,z) \nonumber \\
+ \int d\rr^\prime v(\rho^\prime,z^\prime) V_N(\rr-\rr^\prime) \psi_0(\rho^\prime,z^\prime)e^{im\varphi^\prime}\psi_0(\rho,z) \nonumber \\
\end{eqnarray}
\begin{eqnarray}
\label{BdG2}
-\omega \, v(\rho,z)e^{im\varphi} = \left[ H_0 - \mu \right. \nonumber \\
\left. + \int d\rr^\prime \psi^*_0(\rho^\prime,z^\prime) V_N(\rr-\rr^\prime) \psi_0(\rho^\prime,z^\prime)\right] v(\rho,z)e^{im\varphi} \nonumber \\
+ \int d\rr^\prime \psi_0(\rho^\prime,z^\prime) V_N(\rr-\rr^\prime) v(\rho^\prime,z^\prime)e^{im\varphi^\prime}\psi_0^*(\rho,z) \nonumber \\
+ \int d\rr^\prime u(\rho^\prime,z^\prime) V_N(\rr-\rr^\prime) \psi_0^*(\rho^\prime,z^\prime)e^{im\varphi^\prime}\psi_0^*(\rho,z), \nonumber \\
\end{eqnarray}
where $H_0 = -\frac{\hbar^2}{2M} \nabla^2 + U(\rr)$ and $V_N(\rr-\rr^\prime) = (N-1)V(\rr-\rr^\prime)$.  By using Hankel transforms to compute the interaction terms in Eqs.~(\ref{BdG1}) and~(\ref{BdG2}), we are able to account for the angular dependence of the integrands by using a Bessel function of the appropriate order, as described in Appendix~\ref{app:numerics}.  To solve these equations, we write them in matrix form, as in~\cite{Ronen06a,Lundh06}, and diagonalize them numerically to find the eigenvectors $(u_i,v_i^*)$ and the eigenvalues $\omega_i$.

We point out that while there are solutions of the BdG equations of the form $(u_i,v_i^*)$, there are always solutions of the form $(v_i^*,u_i)$ with $\omega_i \rightarrow -\omega_i$ and $m\rightarrow -m$.  For the case of $s=0$ BECs, there is a solution of the original form $(u_i,v_i^*)$ with the same $\omega_i$ but with $m \rightarrow -m$.  This simply expresses the fact that counter-rotating excitations are degenerate due to the reflection symmetry of the $s=0$ ground state.  The presence of a vortex breaks this double degeneracy.  We shall say that the excitation $(u_i,v_i^*)$ has a positive norm when $\int d\rr \left[ |u(\rho,z)|^2-|v(\rho,z)|^2 \right] > 0$.  It can then be normalized such that $\int d\rr \left[ |u(\rho,z)|^2-|v(\rho,z)|^2 \right]=1$.  The solution $(v_i^*,u_i)$ with $\omega_i \rightarrow -\omega_i$ and $m \rightarrow -m$ will then have a negative norm, obeying $\int d\rr \left[ |u(\rho,z)|^2-|v(\rho,z)|^2 \right]=-1$.  A positive norm mode with a negative energy eigenvalue signifies that there exists a lower energy solution of the GPE; the same situation is represented by a negative norm mode with a positive energy eigenvalue~\cite{Svidzinsky98}.

The solutions of the BdG equations characterize the stability of $s=1$ DBECs.  The global thermodynamical instability of $s=1$ DBECs is seen as a negative norm BdG mode with $m=1$ and positive energy for all trap aspect ratios and dipolar interaction strengths.  This mode corresponds to the system's decay into the energetically favored rotationless ground state, just as for BECs with purely contact interactions.  The component of the mode with angular dependence $e^{-i(m-s)\varphi}=1$ is in this case rotationless, capturing the symmetry of the $s=0$ ground state.  Since this mode populates the core of the vortex, it is refered to as a core mode.   However, at ultracold temperatures, thermodynamical stability is less relevant in characterizing the stability of a condensate since there needs to be some thermal processes acting on the system to dissipitavely drive it into a lower energy state.  We therefore disregard thermodynamical instability in the following.

Instead, we focus on studying the dynamical stability of $s=1$ DBECs.  The emergence of a BdG energy eigenvalue with a nonzero imaginary part corresponds to a dynamical instability in the system~\cite{BEC2003}.  For example, suppose that a BdG mode $(u_i,v^*_i)$ has energy $\omega_i = \omega_R + i \omega_I$ with $\omega_I \neq 0$; then the mode will have the time dependences $e^{(\omega_I-i\omega_R)t}$ and $e^{-(\omega_I - i\omega_R)t}$, either exponentially growing or decaying in time.  Consequently, $\omega_I$ determines the rate of decay of the condensate, given by $\tau = 1/\omega_I$.

\begin{figure}
\includegraphics[width=\columnwidth]{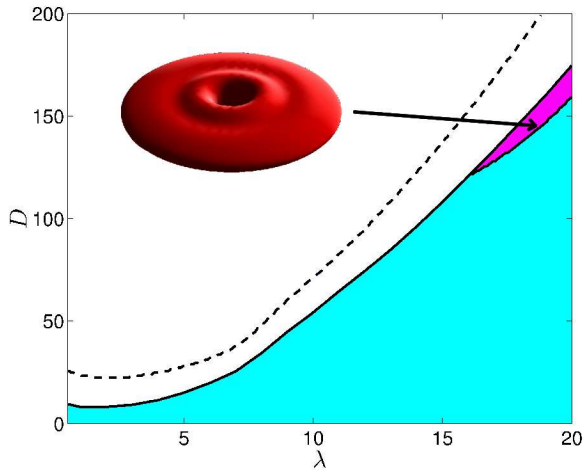}
\caption{\label{fig:stab}  (color online).  The black dashed line marks the maximum dipole strength $D$, for a given trap aspect ratio $\lambda$, above which the Gross-Pitaevskii Energy functional has no minimum corresponding to the $s=1$ DBEC.  The solid line marks a more restrictive stability line, determined by the onset of dynamical instability, signaled by the emergence of an imaginary energy in an excitation mode.  The pink (darker) region represents where oscillations with local minima are observed on dynamically stable states.  The inset is an isodensity surface plot of an $s=1$ DBEC at the point in parameter space indicated by the arrow.  The ripples in the density are explained in Ref~\cite{Wilson08}. }
\end{figure}

We determine where $s=1$ DBECs are dynamically stable by finding the region in parameter space where all of the BdG modes have purely real energy eigenvalues.  This region is shown by the colored portion of Figure~\ref{fig:stab}.  The dashed line in this figure marks, for a given $\lambda$, the $D$ below which we find a local minimum of the GP energy functional~\cite{Ronen06a} by using our reduced 2D algorithm.  We find that, for all $\lambda$, the $D$ above which the GP energy functional has no minimum corresponding to an $s=1$ ground state and the $D$ at which the BdG spectrum begins to possess imaginary energy, denoted $D_\mathrm{crit}$, are never equal.  Indeed, dynamical instability occurs for values of $D$ at which the GPE has a solution.  This is because in the 2D minimization of the vortex-state energy, perturbations that break the $s=1$ angular momentum are not allowed, these are only examined later with the BdG equations.  Using a fully 3D calculation, we check the accuracy of $D_\mathrm{crit}$ for various trap aspect ratios by time evolving the condensate wave function with an initial random perturbation.  The $D_\mathrm{crit}$ that we calculate using the 3D algorithm, corresponding to the $D$ at which we observe collapse under time evolution, agrees with the $D_\mathrm{crit}$ that we calculate by finding imaginary energy eigenvalues in the BdG spectrum using our 2D algorithm.  The pink (darker) region in Figure~\ref{fig:stab} represents the region where we find dynamically stable $s=1$ ground states having radial ripples with local minima, as illustrated by the inset.  This feature has been explained in detail in Ref.~\cite{Wilson08}.

We find that $s=1$ DBECs possess imaginary energy in their BdG spectrum only when two modes of opposite norm are degenerate with each other, just as is the case for a BEC with contact interactions.  This circumstance was recently studied in Ref.~\cite{Nakamura08}, where it is confirmed perturbatively for BECs with contact interactions.  Ref.~\cite{Lundh06} also confirmed this claim using a two-mode approximation.  Indeed, we find that the same holds true for DBECs, where the only difference between the two systems is the shape of the mean field potential.  At all aspect ratios, we observe, for some finite value of $D$, two modes with opposite norm approach and then go degenerate with each other at $D_\mathrm{crit}$.  At the point of degeneracy, the modes develop equal and opposite imaginary energies, signifying dynamical instability.  If two modes that have the same norm approach each other, they undergo an avoided crossing instead of becoming degenerate.

For a BEC with pure contact interactions in the $s=1$ vortex state, the mode that defines the onset of dynamical instability is independent of aspect ratio $\lambda$.  Positive contact interactions ensure dynamical stability while negative contact interactions (for $\lambda \gtrsim 0.3$) bring about a dynamical instability due to an $m=2$ mode~\cite{Saito02}.  Additionally, the $s=2$ state is dynamically unstable due to an $m=2$ mode for negative contact interactions, while an $m=2$ mode defines windows of dynamical stability for positive contact interactions.  This holds true for these systems no matter how oblate the trap.  

\begin{figure}
\includegraphics[width=\columnwidth]{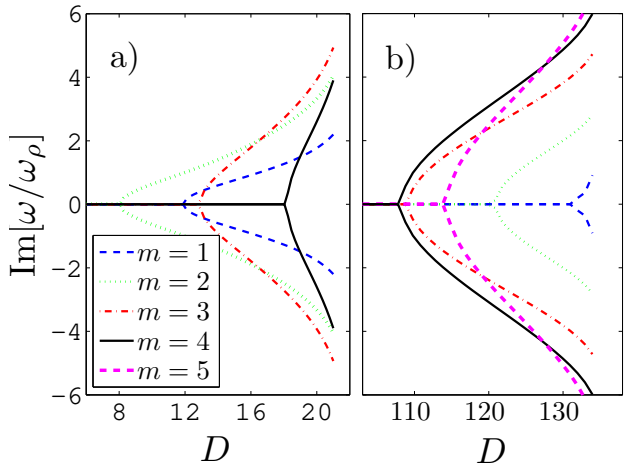}
\caption{\label{fig:exDimag} (color online).  a)  The imaginary part of the BdG excitation spectrum for a number of $m$ values for a DBEC in a trap with aspect ratio a) $\lambda = 2$ and b) $\lambda=15$.  For $\lambda=2$, the $m=2$ modes clearly develop imaginary energy eigenvalues at a $D$ smaller than the other modes, defining $D_\mathrm{crit}\simeq 8$ for this aspect ratio.  For $\lambda=15$, the $m=4$ modes develop imaginary energy at a $D$ smaller than the other modes, defining $D_\mathrm{crit} \simeq 108$ for this aspect ratio.}
\end{figure}

The case for a DBEC, however, is quite different.  Figure~\ref{fig:exDimag} illustrates the imaginary parts of the BdG spectrum for $m=1-5$ for DBECs in traps with aspect ratios $\lambda = 2$ and $\lambda = 15$.  Where these imaginary energies are zero, from $D=0$ to $D_\mathrm{crit}$, the condensates are dynamically stable.  Notice that for $\lambda = 2$, an $m=2$ mode develops imaginary energy at a $D$ well below the other modes, defining $D_\mathrm{crit}$ for this aspect ratio.  However, an $m=4$ mode serves to define $D_\mathrm{crit}$ for $\lambda = 15$.  

\begin{figure}
\includegraphics[width=\columnwidth]{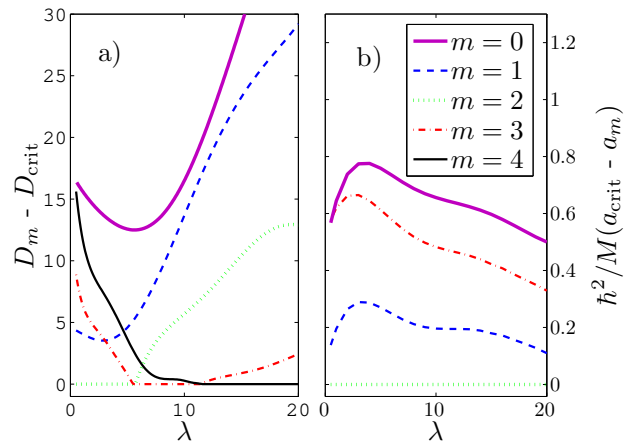}
\caption{\label{fig:exm} (color online).  a)  The difference between the $D$ at which BdG modes with different $m$ quantum numbers first develop imaginary energy eigenvalues and $D_\mathrm{crit}$, the smallest $D$ at which any mode develops an imaginary energy eigenvalue.  Modes with larger $m$ serve to define $D_\mathrm{crit}$ for more oblate traps.  b)  The same, but for negative contact interactions instead of dipole-dipole interactions.  An $m=2$ BdG mode is always the first to develop an imaginary energy eigenvalue for this case, except for trap aspect ratios $\lambda \lesssim 0.3$, for which an $m=1$ mode plays this role.}
\end{figure}

Indeed, unlike BECs with contact interactions, modes with different $m$ quantum numbers serve to define $D_\mathrm{crit}$ at different aspect ratios for DBECs.  For moderate trap aspect ratios (such as $\lambda=2$), an $m=2$ mode defines $D_\mathrm{crit}$ for the DBEC, similar to the case for contact interactions.  However, as the trap aspect ratio is increased to more oblate shapes, modes with larger $m$ quantum numbers develop imaginary energy eigenvalues at smaller values of $D$ than the $m=2$ mode.  Figure~\ref{fig:exm}(a) illustrates this by plotting the \emph{differences} between the $D$'s at which the BdG modes with different angular symmetries first develop imaginary energy eigenvalues, and $D_\mathrm{crit}$, as a function of $\lambda$.  Thus, for a given $\lambda$ the lowest curve identifies the symmetry of the unstable mode.  For $6 \lesssim \lambda \lesssim 12$, an $m=3$ mode defines $D_\mathrm{crit}$ while for larger aspect ratios, an $m=4$ mode defines $D_\mathrm{crit}$.  Although it is not shown here, we find that at even larger aspect ratios the vortex decays into still higher $m$-modes.  

The relevance of the $m$-dependent dynamical instability is that the dipole-dipole interaction leads a BEC to instability locally and at a fixed length scale, the wavelength of which is determined by the axial harmonic oscillator length.  We find that, at the onset of imaginary energy, these modes have radial nodal spacings very similar to that of the roton on the rotationless DBEC, namely $\lambda / 2 \sim \pi l_z$, where $\l_z = \sqrt{\hbar / M \omega_z}$ is the axial harmonic oscillator length.  The angular dependence of these modes behaves in the same way.  Increasing $\lambda$ decreases the ratio $l_z / a_\mathrm{ho}$, so more radial nodes, fixed by $l_z$, can fit into the condensate for larger $\lambda$.  In the same way, more angular nodes can fit into the condensate, therefore bringing about dynamical instability by modes with larger $m$ quantum number, and hence more angular nodes.

All of the previously discussed BdG modes that we identify as being responsible for dynamical instability are axially symmetric and nodeless in $z$.  Modes that break this axial symmetry can correspond to vortex excitations, where the vortex core itself may tilt or bend, and have been termed ``kelvon'' modes.  Ref.~\cite{Klawunn08} reports that, for a singly-quantized vortex in a DBEC that is otherwise spatially homogeneous, the condensate is dynamically unstable to a kelvon mode when an external periodic potential is applied along the direction of the vortex.  We find that, in a harmonically trapped DBEC, a mode with a single node at $z=0$ determines $D_\mathrm{crit}$ for $\lambda \lesssim 0.28$.  Modes of this type might therefore correspond to a kelvon-instability in prolate traps, but we leave these considerations for future work.

As was done in Ref.~\cite{Saito02} for self-attractive BECs in the singly-quantized vortex state, we perform time-dependent simulations of a DBEC where $D$ is chosen to be just above $D_\mathrm{crit}$, enabling us to go beyond the small deviations from the stationary vortex state and see the actual process of collapse.  Initializing the simulations with random noise, we observe collapse, at all aspect ratios, with an angular symmetry corresponding to the $m$ quantum number of the mode that first develops an imaginary energy eigenvalue.

\section{Stability of Vortices with $s=2$}
\label{sec:s2}

The dynamical instability of condensates with doubly-quantized vortices and contact interactions has been studied extensively~\cite{Pu99,Carr06,Mihalache06,Lundh06}.  These studies report windows of positive scattering length where the BECs are dynamically unstable to an $m=2$ BdG core mode, as well as dynamical instability for all values of negative scattering length due to an $m=2$ mode.  Knowing that the dipolar mean field in a DBEC can be engineered to be more attractive or repulsive for smaller or larger trap aspect ratios, respectively, we investigate the presence of these features in $s=2$ DBECs. When the harmonic trap is more spherical, the dipoles are free to stack vertically, creating an attractive mean field in this direction.  However, in pancake shaped traps the dipoles create a more repulsive mean field.  Thus, for larger trap aspect ratios DBECs are more self-repulsive than for smaller aspect ratios, mimicking the mean field of condensates with repulsive contact interactions.  As an example, we calculate the contribution of the dipolar mean field to the energy of a DBEC in a trap with aspect ratio $\lambda = 2$ and with $\lambda = 15$ for a fixed $D=10$.  In the $\lambda = 15$ trap, we find that this contribution is about five times larger than in the $\lambda = 2$ trap.

\begin{figure}
\includegraphics[width=\columnwidth]{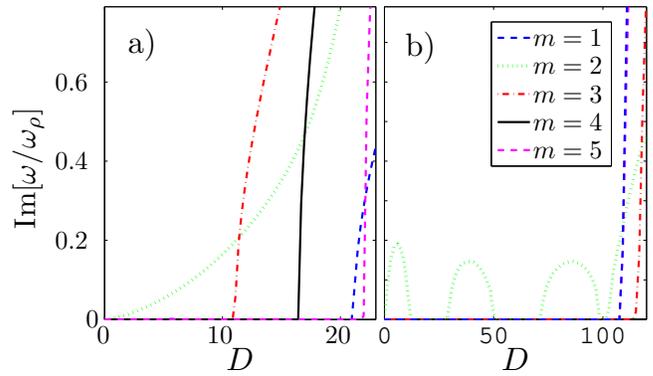}
\caption{\label{fig:k2} (color online).  The positive imaginary part of the energy eigenvalues for a doubly quantized ($s=2$) DBEC in a trap with aspect ratio a) $\lambda=2$ and b) $\lambda=15$.  For $\lambda=2$, an $m=2$ mode possesses imaginary energy for all values of $D$, signifying a dynamical instability for all $D$.  However, for $\lambda=15$ there are windows of dynamical stability as the $m=2$ mode alternates having and not having imaginary energy for different values of $D$, as is the case for a purely repulsive $s=2$ BEC.}
\end{figure}

Indeed, for a DBEC in a trap with aspect ratio $\lambda = 2$ we find that there exists an $m=2$ mode with a complex energy eigenvalue for all values of $D$.  However, for $\lambda = 15$ we find that there are windows in $D$ where an $m=2$ BdG mode has a complex energy eigenvalue, while this same mode has purely real energy outside of these windows, as illustrated in Figure~\ref{fig:k2}.  For trap aspect ratios $\lambda \lesssim 7.5$, there are no windows of dynamical stability and the condensate is dynamically unstable for all $D$.  However, windows of dynamical stability appear for aspect ratios $\lambda \gtrsim 7.5$ and continue for larger $\lambda$.  As is reported in Ref.~\cite{Lundh06}, we find that there is an $m=2$ core mode with negative norm and positive real energy that increases monotonically as it goes successively degenerate with positive norm modes as $D$ is increased to create the windows of dynamical instability.  This mode represents the $s=2$ condensate's instability to splitting into a condensate with two singly quantized vortices.  The core mode is thermodynamically unstable for all values of $D$ and is only dynamically unstable for the windows shown in Figure~\ref{fig:k2}.

\section{Conclusion}

We have implemented a novel 2D algorithm to study the stability and excitations of harmonically trapped DBECs with single vortices by taking advantage of the cylindrical symmetry of the system.  By solving the BdG equations for the $s=1$ DBEC, we systematically map its dynamical stability as a function of trap aspect ratio and dipole-dipole interaction strength $D$.  The BdG spectrum reveals a dynamical instability in the form of a complex energy eigenvalue.  The value of $D$ at which this imaginary energy appears marks the threshold of dynamical stability, $D_\mathrm{crit}$, for the given trap aspect ratio.  By inspecting the BdG spectrum for all $m$ quantum numbers, we determine the symmetry of the mode that is responsible for the dynamical instability in the condensate.  We find, in contrast to BECs with purely contact interactions, that DBECs with a singly-quantized vortex go unstable to modes with larger $m$ quantum numbers for larger trap aspect ratios, signifying a type of local collapse of these condensates.  We have checked the accuracy of $D_\mathrm{crit}$ for various trap aspect ratios by performing fully 3D simulations.  Additionally, we find that, in analogy to a self-repulsive BEC with a doubly-quantized vortex, at larger trap aspect ratios there are successive regions in $D$ where the $s=2$ DBECs are dynamically unstable due to an $m=2$ core mode, while the condensates are dynamically stable outside of these regions.   

\begin{acknowledgments}
The authors acknowledge the financial support of the U.S. Department of Energy and of the National Science Foundation.
\end{acknowledgments}

\appendix

\section{Hankel Transforms and Interpolation}
\label{app:numerics}

Consider the Hankel transform of a function $f(\rho)$, $\tilde{f}(k)$, where $f(\rho) \simeq 0$ for $\rho > R$ and $\tilde{f}(k) \simeq 0$ for $k > K$, and define $S \equiv RK$.  The discrete Hankel transform (DHT) of order $m$ of $f(\rho)$ is then given by~\cite{Guizar04}
\begin{equation}
\label{DHT}
\tilde{f}(k_{mi}) = \frac{2}{K^2}\sum_{j=1}^{N}\frac{f(\rho_{mj})}{J^2_{m+1}(\alpha_{mj})}J_m\left( \frac{\alpha_{mj}\alpha_{mi}}{S} \right)
\end{equation}
where $i,j = 1,\ldots,N$, $\rho_{mj} = \alpha_{mj} / K$, $k_{mi} = \alpha_{mi} / R$ and $\alpha_{mi}$ is the $i^\mathrm{th}$ root of $J_m(\rho)$.  Conversely, the inverse DHT of order $m$ of $\tilde{f}(k)$ is given by
\begin{equation}
\label{iDHT}
f(\rho_{mi}) = \frac{2}{R^2}\sum_{j=1}^{N} \frac{\tilde{f}(k_{mj})}{J^2_{m+1}(\alpha_{mj})} J_m\left( \frac{\alpha_{mj}\alpha_{mi}}{S} \right)_.
\end{equation}
Eqs.~(\ref{DHT}) and~(\ref{iDHT}) show that in order to perform a DHT of order $m$ on a function, the function should be defined on a grid proportional to the zeros of the $m$-order Bessel function, $\rho_{mi}$.

For the problem we are considering, the functions $f(\rho)$ and $\tilde{f}(k)$ have the angular dependence $e^{im\varphi}$ and $e^{-imk_\varphi}$, respectively, which is why the DHTs above are written with Bessel functions of order $m$.  When we calculate the BdG modes for the vortex states of a DBEC (see section~\ref{sec:BdG}), we take DHTs of functions like $\psi_0^*(\rho,z)e^{-is\varphi}u(\rho,z)e^{i(m+s)\varphi}$ where $\psi_0^*(\rho,z)$ is defined on a grid proportional to the zeros of the $s$-order Bessel function and $u(\rho,z)$ is defined on a grid proportional the the zeros of the Bessel function of order $m+s$.  To perform a DHT on the $\rho$-coordinate of the function $\psi_0^*(\rho,z)u(\rho,z)$, the function must be defined on a grid proportional to the zeros of the Bessel function of order $m$, so $\psi_0^*(\rho,z)$ and $u(\rho,z)$ must both be interpolated onto this grid.  To accomplish this, we have developed an accurate interpolation scheme based on the DHT itself.


As explained in Ref.~\cite{ArfkenWeber}, the function $f(\rho)$ may be expanded in an $m^\mathrm{th}$ order Bessel series,
\begin{equation}
\label{B1}
f(\rho) = \sum_{i=1}^N c_{mi}J_m\left( \alpha_{mi}\frac{\rho}{R} \right)
\end{equation}
where the coefficients $c_{mi}$ are given by
\begin{equation}
\label{B2}
c_{mi} = \frac{2}{R^2\,[J_{m+1}(\alpha_{mi})]^2}\int_0^R f(\rho)J_m\left( \alpha_{mi}\frac{\rho}{R} \right)\rho \, d\rho.
\end{equation}
The integral in Eq.~(\ref{B2}) is just the Hankel transform~(\ref{transform}) with $\alpha_{mi}/R = k_{mi}$, giving the transformed function $\tilde{f}(k_i)$.  If $\rho$ is discretized in Eq.~(\ref{B1}), then this prescription gives exactly Eq.~(\ref{iDHT}).

We wish to consider the case where our function is defined on the grid $\rho_{mi}$, proportional to the zeros of the Bessel function of order $m$, but it needs to be defined on the grid $\rho_{ni}$, with $n\neq m$.  To do this, we expand $f(\rho_{ni})$ in a  Bessel series,
\begin{equation}
\label{int1}
f(\rho_{ni}) = \sum_{j=1}^N c_{mj} J_m\left( \alpha_{mj} \frac{\rho_{ni}}{R} \right)
\end{equation}
where the coefficients $c_{mj}$ are given by Eq.~(\ref{B2}) and are computed in terms of the zeros of the Bessel function of order $m$.  However, in Eq.~(\ref{int1}), the function is expanded on the grid $\rho_{ni}$, proportional to the zeros of the Bessel function of order $n$.  The interpolation algorithm then simply follows by inserting the expression for the coefficients,
\begin{equation}
\label{int2}
f(\rho_{ni}) = \frac{2}{R^2} \sum_{j=1}^N \frac{\tilde{f}(k_{mj})}{J^2_{m+1}(\alpha_{mj})} J_m\left( \frac{\alpha_{mj}\alpha_{ni}}{S} \right)_,
\end{equation}
where $\tilde{f}(k_{mj})$ is the discrete Hankel transform of $f(\rho_{mj})$, Eq.~(\ref{DHT}).


\begin{thebibliography}{30}
\expandafter\ifx\csname natexlab\endcsname\relax\def\natexlab#1{#1}\fi
\expandafter\ifx\csname bibnamefont\endcsname\relax
  \def\bibnamefont#1{#1}\fi
\expandafter\ifx\csname bibfnamefont\endcsname\relax
  \def\bibfnamefont#1{#1}\fi
\expandafter\ifx\csname citenamefont\endcsname\relax
  \def\citenamefont#1{#1}\fi
\expandafter\ifx\csname url\endcsname\relax
  \def\url#1{\texttt{#1}}\fi
\expandafter\ifx\csname urlprefix\endcsname\relax\def\urlprefix{URL }\fi
\providecommand{\bibinfo}[2]{#2}
\providecommand{\eprint}[2][]{\url{#2}}

\bibitem[{\citenamefont{Koch et~al.}(2008)\citenamefont{Koch, Lahaye, Metz,
  Frohlich, Griesmaier, and Pfau}}]{KochNature08a}
\bibinfo{author}{\bibfnamefont{T.}~\bibnamefont{Koch}},
  \bibinfo{author}{\bibfnamefont{T.}~\bibnamefont{Lahaye}},
  \bibinfo{author}{\bibfnamefont{J.}~\bibnamefont{Metz}},
  \bibinfo{author}{\bibfnamefont{B.}~\bibnamefont{Frohlich}},
  \bibinfo{author}{\bibfnamefont{A.}~\bibnamefont{Griesmaier}},
  \bibnamefont{and} \bibinfo{author}{\bibfnamefont{T.}~\bibnamefont{Pfau}},
  \bibinfo{journal}{Nature Physics} \textbf{\bibinfo{volume}{4}},
  \bibinfo{pages}{218} (\bibinfo{year}{2008}).

\bibitem[{\citenamefont{Lahaye et~al.}(2008)\citenamefont{Lahaye, Koch,
  Frohlich, Fattori, Metz, Griesmaier, Giovanazzi, and Pfau}}]{Lahaye08}
\bibinfo{author}{\bibfnamefont{T.}~\bibnamefont{Lahaye}},
  \bibinfo{author}{\bibfnamefont{T.}~\bibnamefont{Koch}},
  \bibinfo{author}{\bibfnamefont{B.}~\bibnamefont{Frohlich}},
  \bibinfo{author}{\bibfnamefont{M.}~\bibnamefont{Fattori}},
  \bibinfo{author}{\bibfnamefont{J.}~\bibnamefont{Metz}},
  \bibinfo{author}{\bibfnamefont{A.}~\bibnamefont{Griesmaier}},
  \bibinfo{author}{\bibfnamefont{S.}~\bibnamefont{Giovanazzi}},
  \bibnamefont{and} \bibinfo{author}{\bibfnamefont{T.}~\bibnamefont{Pfau}},
  \bibinfo{journal}{Nature} \textbf{\bibinfo{volume}{448}},
  \bibinfo{pages}{672} (\bibinfo{year}{2008}).

\bibitem[{\citenamefont{Ronen et~al.}(2007)\citenamefont{Ronen, Bortolotti, and
  Bohn}}]{Ronen07}
\bibinfo{author}{\bibfnamefont{S.}~\bibnamefont{Ronen}},
  \bibinfo{author}{\bibfnamefont{D.~C.~E.} \bibnamefont{Bortolotti}},
  \bibnamefont{and} \bibinfo{author}{\bibfnamefont{J.~L.} \bibnamefont{Bohn}},
  \bibinfo{journal}{Phys. Rev. Lett.} \textbf{\bibinfo{volume}{98}},
  \bibinfo{pages}{030406} (\bibinfo{year}{2007}).

\bibitem[{\citenamefont{Baranov}(2008)}]{Baranov08PR}
\bibinfo{author}{\bibfnamefont{M.~A.} \bibnamefont{Baranov}},
  \bibinfo{journal}{Physics Reports} \textbf{\bibinfo{volume}{464}},
  \bibinfo{pages}{71} (\bibinfo{year}{2008}).

\bibitem[{\citenamefont{Wilson et~al.}(2008)\citenamefont{Wilson, Ronen, Pu,
  and Bohn}}]{Wilson08}
\bibinfo{author}{\bibfnamefont{R.~M.} \bibnamefont{Wilson}},
  \bibinfo{author}{\bibfnamefont{S.}~\bibnamefont{Ronen}},
  \bibinfo{author}{\bibfnamefont{H.}~\bibnamefont{Pu}}, \bibnamefont{and}
  \bibinfo{author}{\bibfnamefont{J.~L.} \bibnamefont{Bohn}},
  \bibinfo{journal}{Phys. Rev. Lett.} \textbf{\bibinfo{volume}{100}},
  \bibinfo{pages}{245302} (\bibinfo{year}{2008}).

\bibitem[{\citenamefont{Santos et~al.}(2003)\citenamefont{Santos, Shlyapnikov,
  and Lewenstein}}]{Santos03a}
\bibinfo{author}{\bibfnamefont{L.}~\bibnamefont{Santos}},
  \bibinfo{author}{\bibfnamefont{G.~V.} \bibnamefont{Shlyapnikov}},
  \bibnamefont{and}
  \bibinfo{author}{\bibfnamefont{M.}~\bibnamefont{Lewenstein}},
  \bibinfo{journal}{Phys. Rev. Lett.} \textbf{\bibinfo{volume}{90}},
  \bibinfo{pages}{250403} (\bibinfo{year}{2003}).

\bibitem[{\citenamefont{Kawaguchi et~al.}(2006)\citenamefont{Kawaguchi, Saito,
  and Ueda}}]{Kawaguchi06}
\bibinfo{author}{\bibfnamefont{Y.}~\bibnamefont{Kawaguchi}},
  \bibinfo{author}{\bibfnamefont{H.}~\bibnamefont{Saito}}, \bibnamefont{and}
  \bibinfo{author}{\bibfnamefont{M.}~\bibnamefont{Ueda}},
  \bibinfo{journal}{Phys. Rev. Lett.} \textbf{\bibinfo{volume}{96}},
  \bibinfo{pages}{080405} (\bibinfo{year}{2006}).

\bibitem[{\citenamefont{Vengalattore et~al.}(2008)\citenamefont{Vengalattore,
  Leslie, Guzman, and Stamper-Kurn}}]{Vengalattore08}
\bibinfo{author}{\bibfnamefont{M.}~\bibnamefont{Vengalattore}},
  \bibinfo{author}{\bibfnamefont{S.~R.} \bibnamefont{Leslie}},
  \bibinfo{author}{\bibfnamefont{J.}~\bibnamefont{Guzman}}, \bibnamefont{and}
  \bibinfo{author}{\bibfnamefont{D.~M.} \bibnamefont{Stamper-Kurn}},
  \bibinfo{journal}{Phys. Rev. Lett.} \textbf{\bibinfo{volume}{100}},
  \bibinfo{pages}{170403} (\bibinfo{year}{2008}).

\bibitem[{\citenamefont{Yi and Pu}(2006)}]{Pu06}
\bibinfo{author}{\bibfnamefont{S.}~\bibnamefont{Yi}} \bibnamefont{and}
  \bibinfo{author}{\bibfnamefont{H.}~\bibnamefont{Pu}}, \bibinfo{journal}{Phys.
  Rev. A} \textbf{\bibinfo{volume}{73}}, \bibinfo{pages}{061602(R)}
  (\bibinfo{year}{2006}).

\bibitem[{\citenamefont{van Bijnen et~al.}(2007)\citenamefont{van Bijnen,
  O'Dell, Park, and Martin}}]{Bijnen07}
\bibinfo{author}{\bibfnamefont{R.~M.~W.} \bibnamefont{van Bijnen}},
  \bibinfo{author}{\bibfnamefont{D.~H.~J.} \bibnamefont{O'Dell}},
  \bibinfo{author}{\bibfnamefont{N.~G.} \bibnamefont{Park}}, \bibnamefont{and}
  \bibinfo{author}{\bibfnamefont{A.~M.} \bibnamefont{Martin}},
  \bibinfo{journal}{Physical Review Letters} \textbf{\bibinfo{volume}{98}},
  \bibinfo{pages}{150401} (\bibinfo{year}{2007}).

\bibitem[{\citenamefont{Dodd et~al.}(1996)\citenamefont{Dodd, Edwards,
  Williams, Clark, Holland, Ruprecht, and Burnett}}]{Dodd96}
\bibinfo{author}{\bibfnamefont{R.~J.} \bibnamefont{Dodd}},
  \bibinfo{author}{\bibfnamefont{M.}~\bibnamefont{Edwards}},
  \bibinfo{author}{\bibfnamefont{C.~J.} \bibnamefont{Williams}},
  \bibinfo{author}{\bibfnamefont{C.~W.} \bibnamefont{Clark}},
  \bibinfo{author}{\bibfnamefont{M.~J.} \bibnamefont{Holland}},
  \bibinfo{author}{\bibfnamefont{P.~A.} \bibnamefont{Ruprecht}},
  \bibnamefont{and} \bibinfo{author}{\bibfnamefont{K.}~\bibnamefont{Burnett}},
  \bibinfo{journal}{Phys. Rev. A} \textbf{\bibinfo{volume}{54}},
  \bibinfo{pages}{661} (\bibinfo{year}{1996}).

\bibitem[{\citenamefont{Bradley et~al.}(1997)\citenamefont{Bradley, Sackett,
  and Hulet}}]{Bradley97}
\bibinfo{author}{\bibfnamefont{C.~C.} \bibnamefont{Bradley}},
  \bibinfo{author}{\bibfnamefont{C.~A.} \bibnamefont{Sackett}},
  \bibnamefont{and} \bibinfo{author}{\bibfnamefont{R.~G.} \bibnamefont{Hulet}},
  \bibinfo{journal}{Phys. Rev. Lett.} \textbf{\bibinfo{volume}{78}},
  \bibinfo{pages}{985} (\bibinfo{year}{1997}).

\bibitem[{\citenamefont{Parker et~al.}()\citenamefont{Parker, Ticknor, Martin,
  and O'Dell}}]{Parker08}
\bibinfo{author}{\bibfnamefont{N.~G.} \bibnamefont{Parker}},
  \bibinfo{author}{\bibfnamefont{C.}~\bibnamefont{Ticknor}},
  \bibinfo{author}{\bibfnamefont{A.~M.} \bibnamefont{Martin}},
  \bibnamefont{and} \bibinfo{author}{\bibfnamefont{D.~H.~J.}
  \bibnamefont{O'Dell}}, \bibinfo{note}{arXiv:0810.2028v1, (2008)}.

\bibitem[{\citenamefont{Klawunn et~al.}(2008)\citenamefont{Klawunn, Nath,
  Pedri, and Santos}}]{Klawunn08}
\bibinfo{author}{\bibfnamefont{M.}~\bibnamefont{Klawunn}},
  \bibinfo{author}{\bibfnamefont{R.}~\bibnamefont{Nath}},
  \bibinfo{author}{\bibfnamefont{P.}~\bibnamefont{Pedri}}, \bibnamefont{and}
  \bibinfo{author}{\bibfnamefont{L.}~\bibnamefont{Santos}},
  \bibinfo{journal}{Phys. Rev. Lett.} \textbf{\bibinfo{volume}{100}},
  \bibinfo{pages}{240403} (\bibinfo{year}{2008}).

\bibitem[{\citenamefont{Lundh and Nilsen}(2006)}]{Lundh06}
\bibinfo{author}{\bibfnamefont{E.}~\bibnamefont{Lundh}} \bibnamefont{and}
  \bibinfo{author}{\bibfnamefont{H.~M.} \bibnamefont{Nilsen}},
  \bibinfo{journal}{Phys. Rev. A} \textbf{\bibinfo{volume}{74}},
  \bibinfo{pages}{063620} (\bibinfo{year}{2006}).

\bibitem[{\citenamefont{Pu et~al.}(1999)\citenamefont{Pu, Law, Eberly, and
  Bigelow}}]{Pu99}
\bibinfo{author}{\bibfnamefont{H.}~\bibnamefont{Pu}},
  \bibinfo{author}{\bibfnamefont{C.~K.} \bibnamefont{Law}},
  \bibinfo{author}{\bibfnamefont{J.~H.} \bibnamefont{Eberly}},
  \bibnamefont{and} \bibinfo{author}{\bibfnamefont{N.~P.}
  \bibnamefont{Bigelow}}, \bibinfo{journal}{Phys. Rev. A}
  \textbf{\bibinfo{volume}{59}}, \bibinfo{pages}{1533} (\bibinfo{year}{1999}).

\bibitem[{\citenamefont{Bortolotti et~al.}(2006)\citenamefont{Bortolotti,
  Ronen, Bohn, and Blume}}]{Bortolotti06}
\bibinfo{author}{\bibfnamefont{D.~C.~E.} \bibnamefont{Bortolotti}},
  \bibinfo{author}{\bibfnamefont{S.}~\bibnamefont{Ronen}},
  \bibinfo{author}{\bibfnamefont{J.~L.} \bibnamefont{Bohn}}, \bibnamefont{and}
  \bibinfo{author}{\bibfnamefont{D.}~\bibnamefont{Blume}},
  \bibinfo{journal}{Phys. Rev. Lett.} \textbf{\bibinfo{volume}{97}},
  \bibinfo{pages}{160402} (\bibinfo{year}{2006}).

\bibitem[{\citenamefont{Yi and You}(2000)}]{Yi00}
\bibinfo{author}{\bibfnamefont{S.}~\bibnamefont{Yi}} \bibnamefont{and}
  \bibinfo{author}{\bibfnamefont{L.}~\bibnamefont{You}},
  \bibinfo{journal}{Phys. Rev. A} \textbf{\bibinfo{volume}{61}},
  \bibinfo{pages}{041604(R)} (\bibinfo{year}{2000}).

\bibitem[{\citenamefont{Werner et~al.}(2005)\citenamefont{Werner, Griesmaier,
  Hensler, Simoni, Tiesinga, Stuhler, and Pfau}}]{Werner05}
\bibinfo{author}{\bibfnamefont{J.}~\bibnamefont{Werner}},
  \bibinfo{author}{\bibfnamefont{A.}~\bibnamefont{Griesmaier}},
  \bibinfo{author}{\bibfnamefont{S.}~\bibnamefont{Hensler}},
  \bibinfo{author}{\bibfnamefont{A.}~\bibnamefont{Simoni}},
  \bibinfo{author}{\bibfnamefont{E.}~\bibnamefont{Tiesinga}},
  \bibinfo{author}{\bibfnamefont{J.}~\bibnamefont{Stuhler}}, \bibnamefont{and}
  \bibinfo{author}{\bibfnamefont{T.}~\bibnamefont{Pfau}},
  \bibinfo{journal}{Phys. Rev. Lett.} \textbf{\bibinfo{volume}{94}},
  \bibinfo{pages}{183201} (\bibinfo{year}{2005}).

\bibitem[{\citenamefont{Jackson}(1998)}]{JacksonEM}
\bibinfo{author}{\bibfnamefont{J.~D.} \bibnamefont{Jackson}},
  \emph{\bibinfo{title}{Classical Electrodynamics}} (\bibinfo{publisher}{Wiley,
  New York}, \bibinfo{year}{1998}), \bibinfo{edition}{3rd} ed.

\bibitem[{\citenamefont{G\'oral and Santos}(2003)}]{Goral03}
\bibinfo{author}{\bibfnamefont{K.}~\bibnamefont{G\'oral}} \bibnamefont{and}
  \bibinfo{author}{\bibfnamefont{L.}~\bibnamefont{Santos}},
  \bibinfo{journal}{Phys. Rev. A} \textbf{\bibinfo{volume}{66}},
  \bibinfo{pages}{023613} (\bibinfo{year}{2003}).

\bibitem[{\citenamefont{Pitaevskii and Stringari}(2003)}]{BEC2003}
\bibinfo{author}{\bibfnamefont{L.~P.} \bibnamefont{Pitaevskii}}
  \bibnamefont{and}
  \bibinfo{author}{\bibfnamefont{S.}~\bibnamefont{Stringari}},
  \emph{\bibinfo{title}{Bose-Einstein Condensation}}
  (\bibinfo{publisher}{Oxford University Press, New York},
  \bibinfo{year}{2003}).

\bibitem[{\citenamefont{Ronen et~al.}(2006)\citenamefont{Ronen, Bortolotti, and
  Bohn}}]{Ronen06a}
\bibinfo{author}{\bibfnamefont{S.}~\bibnamefont{Ronen}},
  \bibinfo{author}{\bibfnamefont{D.~C.~E.} \bibnamefont{Bortolotti}},
  \bibnamefont{and} \bibinfo{author}{\bibfnamefont{J.~L.} \bibnamefont{Bohn}},
  \bibinfo{journal}{Phys. Rev. A} \textbf{\bibinfo{volume}{74}},
  \bibinfo{pages}{013623} (\bibinfo{year}{2006}).

\bibitem[{\citenamefont{Svidzinsky and Fetter}(1998)}]{Svidzinsky98}
\bibinfo{author}{\bibfnamefont{A.~A.} \bibnamefont{Svidzinsky}}
  \bibnamefont{and} \bibinfo{author}{\bibfnamefont{A.~L.}
  \bibnamefont{Fetter}}, \bibinfo{journal}{Phys. Rev. A}
  \textbf{\bibinfo{volume}{58}}, \bibinfo{pages}{3168} (\bibinfo{year}{1998}).

\bibitem[{\citenamefont{Nakamura et~al.}(2008)\citenamefont{Nakamura, Mine,
  Okumura, and Yamanaka}}]{Nakamura08}
\bibinfo{author}{\bibfnamefont{Y.}~\bibnamefont{Nakamura}},
  \bibinfo{author}{\bibfnamefont{M.}~\bibnamefont{Mine}},
  \bibinfo{author}{\bibfnamefont{M.}~\bibnamefont{Okumura}}, \bibnamefont{and}
  \bibinfo{author}{\bibfnamefont{Y.}~\bibnamefont{Yamanaka}},
  \bibinfo{journal}{Phys. Rev. A} \textbf{\bibinfo{volume}{77}},
  \bibinfo{pages}{043601} (\bibinfo{year}{2008}).

\bibitem[{\citenamefont{Saito and Ueda}(2002)}]{Saito02}
\bibinfo{author}{\bibfnamefont{H.}~\bibnamefont{Saito}} \bibnamefont{and}
  \bibinfo{author}{\bibfnamefont{M.}~\bibnamefont{Ueda}},
  \bibinfo{journal}{Phys. Rev. Lett.} \textbf{\bibinfo{volume}{89}},
  \bibinfo{pages}{190402} (\bibinfo{year}{2002}).

\bibitem[{\citenamefont{Carr and Clark}(2006)}]{Carr06}
\bibinfo{author}{\bibfnamefont{L.~D.} \bibnamefont{Carr}} \bibnamefont{and}
  \bibinfo{author}{\bibfnamefont{C.~W.} \bibnamefont{Clark}},
  \bibinfo{journal}{Phys. Rev. Lett.} \textbf{\bibinfo{volume}{97}},
  \bibinfo{pages}{010403} (\bibinfo{year}{2006}).

\bibitem[{\citenamefont{Mihalache et~al.}(2006)\citenamefont{Mihalache, Mazilu,
  Malomed, and Lederer}}]{Mihalache06}
\bibinfo{author}{\bibfnamefont{D.}~\bibnamefont{Mihalache}},
  \bibinfo{author}{\bibfnamefont{D.}~\bibnamefont{Mazilu}},
  \bibinfo{author}{\bibfnamefont{B.~A.} \bibnamefont{Malomed}},
  \bibnamefont{and} \bibinfo{author}{\bibfnamefont{F.}~\bibnamefont{Lederer}},
  \bibinfo{journal}{Phys. Rev. A} \textbf{\bibinfo{volume}{73}},
  \bibinfo{pages}{043615} (\bibinfo{year}{2006}).

\bibitem[{\citenamefont{Guizar-Sicairos and Gui\'errez-Vega}(2004)}]{Guizar04}
\bibinfo{author}{\bibfnamefont{M.}~\bibnamefont{Guizar-Sicairos}}
  \bibnamefont{and} \bibinfo{author}{\bibfnamefont{J.~C.}
  \bibnamefont{Gui\'errez-Vega}}, \bibinfo{journal}{J. Opt. Soc. Am. A}
  \textbf{\bibinfo{volume}{21}}, \bibinfo{pages}{53} (\bibinfo{year}{2004}).

\bibitem[{\citenamefont{Arfken and Weber}(2005)}]{ArfkenWeber}
\bibinfo{author}{\bibfnamefont{G.~B.} \bibnamefont{Arfken}} \bibnamefont{and}
  \bibinfo{author}{\bibfnamefont{H.~J.} \bibnamefont{Weber}},
  \emph{\bibinfo{title}{Mathematical Methods for Physicists}}
  (\bibinfo{publisher}{Academic Press, San Diego, CA}, \bibinfo{year}{2005}),
  \bibinfo{edition}{6th} ed.

\end{thebibliography}
\end{document}